\documentclass[10pt]{article}
\usepackage{amssymb,amsmath,amsthm,mathrsfs}
\newcommand{\R}{\mathbb{R}}
\newcommand{\tr}{\text{tr}}

\newcommand{\N}{\mathbb{N}}

\newcommand{\Hilbert}{\mathcal{H}}

\newcommand{\eps}{\varepsilon}

\newtheorem{Lemma}{Lemma}
\newtheorem{Theorem}{Theorem}
\newtheorem{Proposition}{Proposition}
\newtheorem{Corollary}{Corollary}
\newtheorem{Definition}{Definition}
\theoremstyle{definition}
\newtheorem{Remark}{Remark}
\newtheorem{ass}{Assumption}

\begin{document}
%
\title{\textbf{\Large Weakly coupled Schr\"odinger
operators on regular metric trees
}}

\author{
Hynek Kova\v{r}{\'\i}k
\footnote{Also on the leave from 
Nuclear Physics Institute, 
Academy of Sciences,
250\,68 \v{R}e\v{z} near Prague, Czech Republic.
}
}
\date{
\footnotesize
\begin{center}
Institute of Analysis, Dynamics
and Modeling, Universit\"at Stuttgart, \\
PF 80 11 40, D-70569  Stuttgart, Germany.
\\
\vspace{0.2cm}
E-mail:
kovarik@mathematik.uni-stuttgart.de 
\end{center}
}
\maketitle
%
%
\begin{abstract}
\noindent
Spectral properties of the Schr\"odinger operator $
A_{\lambda} = -\Delta +\lambda V$ on regular metric trees are studied. It is
shown that as $\lambda$ goes to zero the behavior of the negative eigenvalues
of $A_{\lambda}$ depends on the global structure of the tree.
\end{abstract}

\vspace{0.2cm}


\vspace{0.2cm}


\section{Introduction}

A rooted metric tree $\Gamma$ consists of 
the set of vertices and the set of edges, i.e.~one dimensional intervals
connecting the vertices, see section \ref{prelim} for details. A Schr\"odinger
operator of the type 
$$
A_{\lambda} = -\Delta +\lambda V\, ,\quad \lambda >0\, , \quad \text{in}\quad
L^2(\Gamma)
$$
is usually defined on a set of functions which satisfy the
so-called Kirchhoff boundary conditions at the vertices of the tree,
see (\ref{kirchhoff}), and a 
suitable condition at its root, which ensures the self-adjointness of 
$A_{\lambda}$. Below we
shall always assume that $V$ is symmetric which means that it 
depends only on the distance from the root of $\Gamma$.

Spectral theory of such operators has
recently attracted a considerable attention, \cite{C,Ex,NS1,NS2,SS,S}.  In
\cite{SS} a detailed asymptotic analysis of the counting function of the
discrete eigenvalues of $A_{\lambda}$ in the limit $\lambda\to\infty$ 
was done for a special class of regular trees whose edges have a constant
length. It was shown, see \cite{SS}, that depending on the decay of $V$  
this asymptotics is either of the Weyl type or it is fully determined by the
behavior of $V$ at infinity.

In this paper we are interested in the spectral behavior of $A_{\lambda}$
in the weak coupling when $\lambda\to 0$. The intuitive
expectation is that the weak coupling behavior of $A_{\lambda}$ should depend
on the rate of the growth of the tree $\Gamma$. This is motivated by the fact
that the weak coupling properties of the operator $A_{\lambda}$ in 
$L^2(\R^n)$ depend strongly on $n$. In particular, it is well known, see
\cite{S}, that for $n=1,2$ the negative
eigenvalues of $A_{\lambda}$ appear for any $\lambda>0$, provided $\int_{\R^n}
V<0$, while for $n\geq 3$ the
negative spectrum of $A_{\lambda}$ remains empty for $\lambda$ small enough.    
Moreover, for $\eps(\lambda)$, 
the lowest eigenvalue of $A_{\lambda}$, the following asymptotic
formulae hold true, \cite{S}: 

\noindent For $n=1$:
$$
\eps(\lambda) \sim -\lambda^2 \, ,\quad \lambda\to 0\, .
$$
For $n=2$:
$$
\eps(\lambda) \sim -e^{-\lambda^{-1}} \, ,\quad \lambda\to 0\, .
$$     
 
In our model, we assign to the tree $\Gamma$ 
a so-called global dimension $d$, see Definition \ref{dim}
below. Roughly speaking, it tells us how fast the number of the branches of
$\Gamma$ grows as a function of the distance from the root. If the latter
grows with the power $d-1$ at infinity, then we say that $d$ is
the global dimension of the tree. We use the notation {\it global} in order to
distinguish $d$ from the local dimension of the tree, which is of course one.
Since $d$ can be in general any real number larger or equal to one, 
it is natural to ask how the weak coupling behavior looks
like for non-integer values of $d$ and what is the condition on $V$ under which
the eigenvalues appear. We will show, see section \ref{weak},
that if $d\in[1,2]$ and $\int_{\Gamma} V <0$, then
$A_{\lambda}$ possesses at least one negative eigenvalue for any $\lambda>0$
and for $\lambda$ small enough this eigenvalue is unique and satisfies
\begin{equation} \label{main-asymp}
 c_1\, \lambda^{\frac{2}{2-d}}\,  \leq\,  |\eps(\lambda)| \, \leq \, c_2\,
 \lambda^{\frac{2}{2-d}}\, , \quad 1\leq d <2\, , 
\end{equation}
for some suitable constants $c_1,c_2$. As expected, the power diverges as
$d$ approaches $2$ from the left. Notice, that our result qualitatively agrees
with the precise asymptotic formula for $\eps(\lambda)$ on branching graphs
with one vertex and finitely many edges, which was found in \cite{Ex}. Such
graphs correspond to $d=1$ in our setting.  

In order to study the operator $A_{\lambda}$ we make use of the
decomposition (\ref{decomp}), see Theorem \ref{NS}, 
which was proved by Naimark and Solomyak
in \cite{NS1,NS2}, see also \cite{C}. In section \ref{aux} we
introduce certain auxiliary operators, whose eigenvalues will give us the
estimate on $\eps(\lambda)$ from above and from below. In order to establish
(\ref{main-asymp}) we find the asymptotics of the lowest eigenvalues of the
auxiliary operators, which are of the same order. This is done in section
\ref{short}.  
In section \ref{virtual} it is shown, under some regularity conditions on $V$, 
that if the tree grows too fast, i.e.~$d>2$,
then $A_{\lambda}$ has no weakly coupled eigenvalues at all, that is 
the discrete spectrum remains empty for $\lambda$ small enough.
Besides, in section \ref{number} we 
give some estimates on the number of eigenvalues of the individual
operators in the decomposition (\ref{decomp}), which are used in the proofs of
the main results, but might be of an independent interest as well.

Throughout the text we will employ the notation
$\alpha:= d-1$ and  $\nu : = \frac{2-d}{2}$.
For a real-valued function $f$ and a real non-integer number $\mu$ we will
use the shorthand
$$
f^{\mu} := {\rm sign\, } f\, |f|^{\mu} = \frac{f\, |f|^{\mu}}{|f|}\, .
$$
Finally, given a self-adjoint operator $T$ on a Hilbert space $\Hilbert$ we
denote by
$N_-(T;s)$ the number of eigenvalues, taking into account their multiplicities, 
of $T$ on the left of the point $s$. For
$s=0$ we will write $N_-(T)$ instead of $N_-(T;0)$.

\section{Preliminaries}
\label{prelim}
We define a metric tree $\Gamma$ with the root $o$ following the construction
given in \cite{NS1}. 
Let $\mathcal{V}(\Gamma)$ be the set of vertices and
$\mathcal{E}(\Gamma)$ be the set of edges of $\Gamma$. The distance
$\rho(y,z)$ between any two points 
$y,z\in\Gamma$ is defined in a natural way as the length of the unique path
connecting $y$ and $z$. Consequently, $|y|$ is equal to
$\rho(y,o)$. We write $y\preceq z$ if $y$ lies on the unique simple path
connecting $o$ with $z$. For $y\preceq z$ we define
$$
<y,z> := \{x\in\Gamma\, :\, y\preceq x\preceq z\}\, .
$$
If $e= <y,z>$ is an edge, then $y$ and $z$ are its endpoints.
For any vertex $z$ its generation $Gen(z)$ is defined by
$$
  Gen(z) =  \#\{x\in \mathcal{V}\, :\,o\prec x\preceq z\}\, .
$$
The branching number $b(z)$ of the vertex $z$ is equal to the number of
edges emanating from $z$. We assume that $b(z) >1$ for any $z\neq o$ and
$b(o)=1$.

\begin{Definition}
A tree $\Gamma$ is called {\rm regular} if all the vertices of the same
generation have equal branching numbers and all the edges of the same
generation have equal length.
\end{Definition}

\noindent
We denote by
$t_k>0$ the distance between the root and the vertices of the $k-$th
generation and by $b_k\in\N$ their corresponding branching number.
For each $k\in\N$ we define the so-called branching function
$g_k:\R_+\to\R_+$ by
\begin{equation*}
g_k(t) := \left\{
\begin{array}{l@{\quad \mathrm{if} \quad }l}
 0 & t < t_k\, , \\
 1 & t_k \leq t \leq  t_{k+1}\, , \\
 b_{k+1}b_{k+2}\cdots b_n & t_n \leq t\leq t_{n+1},\, k<n \, ,
\end{array}
\right.
\end{equation*}
and
$$
g_0(t) := b_0\, b_1\cdots b_n\, \quad t_n \leq t\leq t_{n+1}\, .
$$
It follows directly from the definition that
$$
g_0(t) = \#\{x\in \Gamma\, : \, |x|=t\}\, .
$$
Obviously $g_0(\cdot)$ is a non-decreasing function and the rate of growth of
$g_0$ determines the rate of growth of the tree $\Gamma$. In particular, if
one denotes by $\Gamma(t):= \{x\in\Gamma\, :\, |x|\leq t\}$, the ``ball'' of
radius $t$, then $g_0$ tells us how fast the surface of $\Gamma(t)$ grows
with $t$. This motivates the following

\begin{Definition} \label{dim}
If there exist positive constants $a^-\, ,$ $a^+$ and $T_0$, such that for
all $t\geq T_0$ the inequalities
$$
a^- \leq \frac{g_0(t)}{t^{d-1}} \,  \leq a^+
$$
hold true, then we say that ${\rm d}$ is the {\rm global dimension} of the
tree $\Gamma$.
\end{Definition}

We note that in the case of the so-called homogeneous metric trees treated
in \cite{SS} the function $g_0(t)$ grows faster than any power of $t$.
Formally, this corresponds to $d=\infty$ in the above definition. From now on
we will work under the assumption that $d<\infty$.

\section{Schr\"odinger operators on $\Gamma$}

We will consider potential functions $V$ which satisfy the

\begin{ass} \label{firstassumption}
$V:\R_+\to\R$ is measurable, bounded and $\lim_{t\to\infty}\, V(t)= 0$.
\end{ass}

\noindent For a given function  $V$ which satisfies the Assumption
\ref{firstassumption}
we define the Schr\"odinger operator $A_{\lambda}$ as the self-adjoint
operator in
$L^2(\Gamma)$ associated with the closed quadratic form
$$
Q_{\lambda}[u] := \int_{\Gamma}\left(|u'|^2+\lambda V(|x|)\, |u|^2\right)\, dx
\, ,
$$
with the form domain $D(Q) = H^1(\Gamma)$ consisting of all
continuous functions $u$ such that $u\in H^1(e)$ on each edge $e\in
\mathcal{E}(\Gamma)$ and
$$
\int_{\Gamma}\left(|u'|^2+|u|^2\right)\, dx < \infty\, .
$$
The domain of $A_{\lambda}$ consists of all continuous functions $u$ such that
$u'(o)=0\, $, $u\in H^2(e)$ for each $e\in \mathcal{E}(\Gamma)$ and such that
at each vertex $z\in
\mathcal{V}(\Gamma)\setminus\{o\}$
the matching conditions
\begin{equation} \label{kirchhoff}
u_-(z)= u_1(z)=\cdots = u_{b(z)}(z)\, , \quad  u'_1(z) + \cdots + u'_{b(z)}(z)
= u'_-(z)
\end{equation}
are satisfied, where $u_-$ denotes the restriction of $u$ on the edge
terminating in $z$
and $u_j,\, j=1,...,b(z)$ denote respectively the restrictions of $u$ on the
edges emanating from $z$, see \cite{NS1} for details. Notice that
$A_{\lambda}$ satisfies the Neumann boundary condition at the root $o$. 

The following result by Naimark and Solomyak, see \cite{NS1,NS2}, also
established by Carlson in \cite{C}, makes it possible to reduce the spectral
analysis of  $A_{\lambda}$ to the analysis of one dimensional Schr\"odinger
operators in weighted $L^2(\R_+)$ spaces:

\vspace{0.2cm}

\begin{Theorem} \label{NS}
Let $V$ be
measurable and bounded and suppose that $\Gamma$ is regular. 
Then $A_{\lambda}$ is unitarily equivalent
to the following orthogonal sum of operators:
\begin{equation} \label{decomp}
A_{\lambda} \sim A_{\lambda,0} \oplus \sum_{k=1}^{\infty}\oplus \, A_{\lambda,k}
^{[b_1...b_{k-1}(b_k-1)]}\, .
\end{equation}
\label{NS}
\end{Theorem}
\noindent Here the symbol $ A_{\lambda,k} ^{[b_1...b_{k-1}(b_k-1)]}$
means that the operator $A_{\lambda,k}$ enters the orthogonal sum
$[b_1...b_{k-1}(b_k-1)]$ times. For each $k\in\N$ the corresponding
self-adjoint operator $A_{\lambda,k}$ acts in $L^2((t_k,\infty),\,
g_k)$ and is associated with the closed quadratic form
$$
Q_k[f] = \int_{t_k}^{\infty}\left(|f'|^2+\lambda V(t)\, |f|^2\right)\,
g_k(t)\, dt\, ,
$$
whose form domain is given by the the weighted Sobolev space
$D(Q_k)= H^1_0((t_k,\infty),\, g_k)$ which consists of all functions $f$ such
that 
$$ \int_{t_k}^{\infty}\, \left(|f'|^2+|f|^2\right)\, g_k(t)\, dt
<\infty\, ,\quad  f(t_k)=0\, .
$$
The operator $A_{\lambda,0}$ acts in the weighted space 
$L^2(\R_+,g_0)$ and is associated with the closed form
$$
Q_0[f] = \int_{0}^{\infty}\left(|f'|^2+\lambda V(t)\, |f|^2\right)\, g_0(t)\,
dt\, ,
$$
with the form domain $D(Q_0)=H^1(\R_+,\, g_0)$ which consists of all functions
$f$ such that  
$$
\int_0^{\infty}\, \left(|f'|^2+|f|^2\right)\, g_0(t)\, dt
<\infty \, ,
$$ 
see also \cite{S}.

\subsection{Auxiliary operators}
\label{aux}

Let $d$ be the global dimension of $\Gamma$.
Definition (\ref{dim}) implies that for each $k\in\N_0$
there exist positive constants $a^-_k$ and $a^+_k$, such that
\begin{equation} \label{gkestim}
a^-_k\, (1+t)^{\alpha}=:g^-_k(t)\, \leq \, g_k(t) \leq g^+_k(t):= a^+_k
(1+t)^{\alpha}\,
,\quad t\in [t_k,\infty) \, .
\end{equation}
Now assume that the Rayleigh quotient
$$
\frac{\int_{t_k}^{\infty}\left(|f'|^2+\lambda V(t)\, |f|^2\right)\, g_k(t)\,
  dt}{\int_{t_k}^{\infty}|f|^2\, g_k(t)\, dt}
$$
of the operator $A_{\lambda,k},\, k\geq 0$ is negative for some
$f\in D(Q_k)$. From (\ref{gkestim}) follows that
\begin{eqnarray} \label{Rayleigh}
&& \frac{\int_{t_k}^{\infty}\left(|f'|^2+ \lambda V^-_k(t)\,
    |f|^2\right)\, (1+t)^{\alpha}\,
  dt}{\int_{t_k}^{\infty}|f|^2\, (1+t)^{\alpha}\, dt}  \leq
\frac{\int_{t_k}^{\infty}\left(|f'|^2+\lambda V(t)\, |f|^2\right)\, g_k(t)\,
  dt}{\int_{t_k}^{\infty}|f|^2\, g_k(t)\, dt}  \nonumber  \\
& & \leq
\frac{\int_{t_k}^{\infty}\left(|f'|^2+\lambda V^+_k(t)\,
    |f|^2\right)\, (1+t)^{\alpha}\,
  dt}{\int_{t_k}^{\infty}|f|^2\, (1+t)^{\alpha}\, dt} \, ,
\end{eqnarray}
where
$$
V^-_k(t):= \frac{g_k(t)}{g^-_k(t)}\, V(t)\, , \quad V^+_k(t):=
\frac{g_k(t)}{g^+_k(t)}\, V(t)\, .
$$
It is thus natural to introduce the auxiliary operators $A^{\pm}_{\lambda,k}$
acting in the Hilbert space $L^2((t_k,\infty),\, (1+t)^{\alpha})$ and
associated with the quadratic forms
\begin{equation}
Q^{\pm}_k[f] = \int_{t_k}^{\infty}\left(|f'|^2\,
  +\lambda V^{\pm}_k(t)\, |f|^2 \, \right)\,
(1+t)^{\alpha}\, dt\, , \quad f\in D(Q_k)\, , k\in\N_0\, .
\end{equation}
The variational principle, see e.g. \cite{BS2}, and (\ref{Rayleigh}) 
thus imply that
\begin{equation} \label{numbers}
N_-(A^{+}_{\lambda,k};s)\, \leq \, N_-(A_{\lambda,k};s)\, \leq \,
N_-(A^{-}_{\lambda,k};s) \, ,\quad s\leq 0,\, k\in\N_0\, .
\end{equation}

\noindent Let $E_{n,k}(\lambda)$ be the non-decreasing sequence of negative
eigenvalues of the operators $A_{\lambda,k}$ and let
$E^{\pm}_{n,k}(\lambda)$ be the analogous sequences corresponding to
the operators $A^{\pm}_{\lambda,k}$ respectively. In all these
sequences each eigenvalue occurs according to its multiplicity.
Relation (\ref{numbers}) and variational principle then yield
\begin{equation} \label{eigenvalues}
E^{-}_{n,k}(\lambda)\, \leq \, E_{n,k}(\lambda) \, \leq \, 
E^{+}_{n,k}(\lambda)\, ,\quad
k\in\N_0\, , n\in\N\, ,
\end{equation}
and
\begin{equation} \label{ess-spectra}
\inf \sigma_{ess} (A^{-}_{\lambda,k}) \, \leq \, \inf \sigma_{ess}
(A_{\lambda,k})\,  \leq \, \inf \sigma_{ess} (A^{+}_{\lambda,k})\,
,\quad k\in\N_0\,
\end{equation}

\noindent Next we introduce the transformation $U$ by
$$
(U\, f)(t)  = (1+t)^{\alpha/2}\, f(t) =:\varphi(t)\, ,
$$
which maps
$L^2((t_k,\infty),(1+t)^{\alpha})$ unitarily onto $L^2((t_k,\infty))$.
We thus get

\begin{Lemma} \label{auxiliary}
Let $V$ satisfy the assumptions of Theorem \ref{NS}. Then
\begin{itemize}
\item[(i)] For  each $k\in\N$ the operators $A^{\pm}_{\lambda,k}$ are
unitarily equivalent
to the self-adjoint 
operators $B^{\pm}_{\lambda,k}$ in $L^2((t_k,\infty))$, which act as
\begin{equation}
\left(B^{\pm}_{\lambda,k}\, \varphi\right)(t) = -\varphi''(t)
+\frac{(d-1)(d-3)}{4(1+t)^2}\, \varphi(t)
+\lambda V^{\pm}_k(t)\, \varphi(t)\, ,
\end{equation}
and whose domains consist of all functions $\varphi\in H^2((t_k,\infty))$
such that
$$
\varphi(t_k)=0 \, .
$$
\item[(ii)] $A^{\pm}_{\lambda,0}$
are unitarily equivalent to the self-adjoint operators
$B^{\pm}_{\lambda,0}$ in $L^2(\R_+)$, acting as
\begin{equation}
\left(B^{\pm}_{\lambda,0}\, \varphi\right)(t) = -\varphi''(t)
+\frac{(d-1)(d-3)}{4(1+t)^2}\, \varphi(t)
+\lambda\, V^{\pm}_0(t)\, \varphi(t)\, ,
\end{equation}
with the domain that consists of all $\varphi\in H^2(\R_+)$ such that
\begin{equation} \label{bc}
\varphi'(0) = \frac{d-1}{2}\, \varphi(0)\, .
\end{equation}

\end{itemize}
\end{Lemma}

\begin{proof}
For each $k\in\N_0$ we have
$$
B^{\pm}_{\lambda,k} = U\, A^{\pm}_{\lambda,k}\, U^{-1}\, ,\quad
\|f\|_{L^2((t_k,\infty),\, (1+t)^{\alpha})}\, = \|U f\|_{L^2((t_k,\infty))}\, .
$$
The statement of the Lemma then follows by a direct calculation keeping in
mind that the functions $f$ from the domain of the operators $
A^{\pm}_{\lambda,0}$ satisfy $f'(0)=0$.

\end{proof}

\begin{Remark}
If $V$ satisfies assumption \ref{firstassumption}, then the
inequalities (\ref{ess-spectra}) and standard arguments from the spectral
theory of Schr\"odinger operators, see e.g. \cite[Chap.13.4]{RS4}, imply that
$$
\inf\, \sigma_{ess} (A^{-}_{\lambda,k}) = \inf\, \sigma_{ess}
(A_{\lambda,k}) = \inf\, \sigma_{ess} (A^{+}_{\lambda,k}) = 0\,
,\quad \forall\, k\in\N_0\, .
$$
Moreover, constructing suitable Weyl sequences for the operators
$A_{\lambda,k}$ in the similar way as it was done in \cite{S} for the Laplace
operator, one can actually show that
\begin{equation} \label{essential-k}
\sigma_{ess} (A_{\lambda,k}) = [0,\infty)\, ,\quad  \forall\, k\in\N_0\, .
\end{equation}

\end{Remark}

\section{Number of bound states}
\label{number}

From Theorem \ref{NS} and equation (\ref{essential-k}) we can see that if $V$
satisfies assumption $A$ then
\begin{equation} \label{essential}
\sigma_{ess}\,( A_{\lambda}) = [0,\infty)\, .
\end{equation}
In order to analyze the discrete spectrum of $A_{\lambda}$ we first
study the number of bound states of the individual operators in the
decomposition (\ref{decomp}).

We start by proving an auxiliary Proposition. Given a real valued measurable
bounded function $\tilde{V}$ we consider the  
self-adjoint operator $\tilde{B}_{\lambda}$ acting in $L^2(\R_+)$ as
\begin{equation}
\left(\tilde{B}_{\lambda}\, \varphi\right)(t) = -\varphi''(t)
+\frac{(d-1)(d-3)}{4\, t^2}\, \varphi(t)
+\lambda \tilde{V}(t)\, \varphi(t)\, ,
\end{equation}
and whose domain consist of all functions $\varphi\in H^2(\R_+)$
such that $ \varphi(0)=0$. We have 

\begin{Proposition} \label{Prop1}
Let $d\in[1,2)$. Assume that $\tilde{V}$ satisfies assumption
\ref{firstassumption} and that
$
\int_0^{\infty}\, t\,|\tilde{V}(t)|\, dt <\infty
.
$
Then
\begin{equation} \label{Btilde}
N_-(\tilde{B}_{\lambda}) \leq \lambda\, \tilde{K}(d)\,
\int_0^{\infty}\, t\, |\tilde{V}(t)|\, dt\, ,
\end{equation}
where
$$
\tilde{K}(d) =\frac{\pi}{2\sin(\nu\pi)\Gamma(1-\nu)\Gamma(1+\nu)}\,
.
$$
\end{Proposition}

\begin{proof}
We write
$$
\tilde{B}_{\lambda,0} = \tilde{B}_0 +\lambda\, \tilde{V}\, , \quad
\tilde{B}_0:= -\frac{d^2}{d\, t^2}\, +
\frac{(d-1)(d-3)}{4t^2}\, .
$$
Moreover, without loss of generality we may assume that $\tilde{V}<0$.

By the Birman-Schwinger principle, see e.g.\cite{BS2}, the number of
eigenvalues of $\tilde{B}_{\lambda}$ to the left of the point
$-\kappa^2$ then does not exceed the trace of the operator
$$
\lambda|\tilde{V}|^{1/2}\, (\tilde{B}_0+\kappa^2)^{-1}\, |\tilde{V}|^{1/2}\, .
$$
The integral kernel $\tilde{G}(t,t',\kappa)$ of the operator
$(\tilde{B}_0+\kappa^2)^{-1}$ can be calculated by using the
Sturm-Liouville theory. We get
\begin{equation} 
\tilde{G}(t,t',\kappa) = \left\{
\begin{array}{c@{\quad  }l}
 \frac{\pi\, i}{4}\,  v_1(t,\kappa)\, v_2(t',\kappa)
 & t \geq t' \\
           &                            \\
\frac{\pi\, i}{4}\, v_1(t',\kappa)\, v_2(t,\kappa)   
 &  t < t'
\end{array}
\right. \, ,
\end{equation}
with
\begin{eqnarray*}
v_1(t,\kappa) & = & \sqrt{t}\,\, H_{\nu}^{(1)}(i\kappa t)\, , \\
v_2(t,\kappa) & = & \sqrt{t}\,
H_{\nu}^{(1)}(i\kappa t)+\sqrt{t}\, H_{\nu}^{(2)}(i\kappa t)
\, ,
\end{eqnarray*}
where $H_{\nu}^{(1)}$ resp.~$H_{\nu}^{(2)}$ denote Hankel's functions of
the first resp.~second kind, see e.g. \cite{W}.
Since $\int_0^{\infty}\, t\,
|\tilde{V}(t)|\, dt < \infty$, we can pass to the limit $\kappa\to
0$ in the corresponding integral, using the Lebesgue dominated
convergence theorem, and calculate the trace to get
\begin{equation}
N_-(\tilde{B}_{\lambda}) \leq  \lambda\, \int_0^{\infty}\,
|\tilde{V}(t)|\, |\tilde{G}(t,t,0)| \, dt =  \lambda\,
\tilde{K}(d)\, \int_0^{\infty}\, t\, |\tilde{V}(t)|\, dt\, .
\end{equation}
Here we have used the fact that $\tilde{G}(t,t,\kappa)\to t\, \tilde{K}(d)$ 
pointwise as $\kappa\to 0$, which follows from the asymptotic behavior of the
Hankel functions at zero, see e.g. \cite{AS}. 
\end{proof}

\begin{Remark}
For $d=1$ we have $\tilde{K}(1)=1$ and (\ref{Btilde}) gives the well known
Bargmann inequality, \cite{Ba}.
On the other hand, $\tilde{K}(d)$ diverges as $d\to 2-$.
This is expected because the operator $-\frac{d^2}{d\, t^2}\, -
\frac{1}{4t^2} +\lambda V$ with Dirichlet b.c. at zero does have at
least one negative eigenvalue for any $\lambda >0$ if the integral
of $V$ is negative.
\end{Remark}

\noindent Armed with Proposition \ref{Prop1} we can prove

\begin{Corollary} \label{Cor1}
Let $1\leq d <2$. Assume that $V$ satisfies assumption
\ref{firstassumption} and that
$
\int_0^{\infty}\, t\,|V(t)|\, dt <\infty .
$
Then
\begin{equation} \label{Bargman}
N_-(A_{\lambda,0})  \leq  1 + \lambda\, K(d)\, \int_0^{\infty}\,
|V(t)|\, g_0(t)\, t^{2-d} dt .
\end{equation}
\end{Corollary}

\begin{proof}
We introduce the operator $A^D_{\lambda,0}$, which is associated 
with the quadratic form 
$$
Q^D_0[f] := \int_{0}^{\infty}\left(|f'|^2+\lambda V(t)\, |f|^2\right)\,
g_0(t)\, dt\, , \quad D(Q_0^D)= H^1_0(\R_+,\, g_0)\, ,
$$
where $H^1_0(\R_+,\, g_0):= \{f\in H^1(\R_+,\, g_0),\, f(0)=0\}$.
First we observe that
$$
a\, t^{d-1} \leq g_0(t)\, ,\quad t\in\R_+
$$
for a suitable $a>0$.
We can thus mimic the analysis of Section \ref{aux} and
define the operator
$\tilde{A}_{\lambda}$ acting in $L^2(\R_+,\, t^{d-1})$ associated with the
quadratic form
\begin{equation}
\tilde{Q}[f] = \int_{0}^{\infty}\left(|f'|^2\,
  +\lambda \tilde{V}(t)\, |f|^2 \, \right)\,
t^{d-1}\, dt\, , \quad f\in D(Q)\,
\end{equation}
where $D(Q)=H^1_0((\R_+),\, t^{d-1})$ and
$
\tilde{V}(t):= \frac{g_0(t)}{a\, t^{d-1}}\, V(t)\, .
$
Repeating the arguments of Section \ref{aux} we claim that
$$
N_-(A^D_{\lambda,0}) \leq N_-(\tilde{A}_{\lambda})
$$
and that $\tilde{A}_{\lambda}$ is unitarily equivalent to
$\tilde{B}_{\lambda}$ by means of the transformation $\tilde{U}\, f(t)  =
t^{(d-1)/2}\, f(t)$, which maps $L^2(\R_+,\, t^{d-1})$ unitarily onto 
$L^2(\R_+)$. 
Since the co-dimension of $H^1_0(\R_+,\, g_0)$ in $H^1(\R_+,\, g_0)$ is equal
to one, the variational principle gives 
$$
N_-(A_{\lambda,0})\leq 1+ N_-(A^D_{\lambda,0}) \leq 1+
N_-(\tilde{A}_{\lambda}) = 1 + N_-(\tilde{B}_{\lambda})\, .
$$
Application of Proposition \ref{Prop1} with $\tilde{K}(d)= a\, K(d)$
concludes the proof.
\end{proof}

\begin{Corollary} \label{Cor2}
Let $1\leq d <2$. Let $V$ satisfy assumption \ref{firstassumption} and assume
that
$
\int_0^{\infty}\, t\, |V(t)|\, dt < \infty\, .
$
Then there exists $\lambda_c>0$, so that for $\lambda\in[0,\lambda_c]$ the
discrete spectra of the operators $A_{\lambda,k},\, k\geq 1$ are empty. In
particular we have
\begin{equation} \label{discrete}
\sigma_{d}\, \left(A_{\lambda}\right) = \sigma_{d}\,
\left(A_{\lambda,0}\right)\, ,\quad 0\leq \lambda\leq\lambda_c,
\end{equation}
where the multiplicities of the eigenvalues are taken into account.
\end{Corollary}
\begin{proof}
Let $k\geq 1$ be fixed. In view of Lemma \ref{auxiliary}
it suffices to show that the discrete spectrum of the
operator $B_{\lambda,k}^-$ is empty provided $\lambda$ is small enough.
Since $(d-1)(d-3)\leq 0$, the following inequality holds true in the
sense of quadratic forms:
$$
B_{\lambda,k}^- \geq \mathcal{B}_{\lambda,k} :=
-\frac{d^2}{d t^2} +\frac{(d-1)(d-3)}{4(t-t_k)^2}\, +
\lambda V_k^-(t)\, ,
$$
where the domain of $\mathcal{B}_{\lambda,k}$ coincides with that of
$B_{\lambda,k} ^-$ given in Lemma \ref{auxiliary}. A simple translation $s=
t-t_k$ then shows that $\mathcal{B}_{\lambda,k}$ is unitarily equivalent to
the operator
$$
-\frac{d^2}{d s^2} +\frac{(d-1)(d-3)}{4s^2}\, +
\lambda V_k^-(s+t_k)\,
$$
acting in $L^2(\R_+)$ with Dirichlet boundary condition at zero.
To finish the proof it thus remains to apply Proposition \ref{Prop1} with
$$
\tilde{V}(s) = \frac{g_k(s+t_k)}{g_k^-(s+t_k)}\, V(s+t_k)\, ,\quad
s\in[0,\infty)
$$
and take $\lambda$ small enough, such that $N_-(\mathcal{B}_{\lambda,k})=0$.
\end{proof}

\section{Weak coupling}
\label{weak}

\subsection{The case $1\leq d<2$}
\label{short}

In this section we will show that if $d\in [1,2)$ and $V$ is
attractive in certain
sense, then the operator $A_{\lambda}$ possesses at least one negative
eigenvalue for any $\lambda>0$. Since for
small values of $\lambda$ the discrete spectra of $A_{\lambda}$ and
$A_{\lambda,0}$ coincide,
see Corollary \ref{Cor2}, we will focus on the operator $A_{\lambda,0}$ only.
More exactly, in view of (\ref{eigenvalues}), we will study the operators
$B^{\pm}_{\lambda,0}$. Clearly we have
$$
B^{\pm}_{\lambda,0} = B_0 +\lambda\, V_0^{\pm}\, , \quad
B_0:= -\frac{d^2}{d\, t^2}\, +
\frac{(d-1)(d-3)}{4(1+t)^2}\, ,
$$
with the boundary condition $v'(0)= \frac{d-1}{2}\, v(0)$. Note that,
by Lemma \ref{auxiliary}, the operator $B_0$ is non-negative.
We shall first calculate the Green function of $B_0$ at a point
$-\kappa^2,\, \kappa>0,$ using the Sturm-Liouville theory again. 
In the same manner as in the previous section we obtain 
\begin{equation} \label{greenfunction}
G(t,t',\kappa) := \left\{
\begin{array}{c@{\quad  }l}
 \frac{\pi}{4i \beta(\kappa)}\, \, v_1(t,\kappa)\, v_2(t',\kappa)
 & t \geq t' \\
           &                            \\
\frac{\pi}{4i \beta(\kappa)}\, \, v_1(t',\kappa)\, v_2(t,\kappa)\,
 &  t < t'
\end{array}
\right. \, ,
\end{equation}
where
\begin{eqnarray*}
v_1(t,\kappa) & = & \sqrt{1+t}\,\, H_{\nu}^{(1)}(i\kappa(1+t))\, , \\
v_2(t,\kappa) & = & \sqrt{1+t}\,
\left(H_{\nu}^{(1)}(i\kappa(1+t))-\beta(\kappa)\, H_{\nu}^{(2)}(i\kappa(1+t))
\right)\, , \\
\beta(\kappa) & = & \frac{H_{\nu-1}^{(1)}(i\kappa)}{
  H_{\nu-1}^{(2)}(i\kappa)}\, .
\end{eqnarray*}

Consider a function $W$ which satisfies assumption \ref{firstassumption}.
According to the Birman-Schwinger principle the operator $B_0+\lambda
W$ has an eigenvalue $-\kappa^2$ if and only if the operator
$$
K(\kappa) := |W|^{1/2}\, (B_0+\kappa^2)^{-1}\, W^{1/2}
$$
has eigenvalue $-\lambda^{-1}$. The integral kernel of $K(\kappa)$
is equal to
$$
K(t,t',\kappa) = |W(t)|^{1/2}\, G(t,t',\kappa)\,( W(t'))^{1/2}\, .
$$
We will use the decomposition
$$
K(t,t',\kappa) = L(t,t',\kappa) +M(t,t',\kappa)\, ,
$$
with
$$
L(t,t',\kappa) := \frac{\pi\, 2^{2\nu-1}\, \kappa^{-2\nu}}{(\Gamma(1-\nu))^2\,
  \sin(\nu\pi)}\, \, |W(t)|^{1/2}\, [(1+t)(1+t')]^{-\nu+\frac 12}\,
W(t')^{1/2}\, ,
$$
and denote by $L(\kappa)$ and $M(\kappa)$ the integral operators
with the kernels $L(t,t',\kappa)$ and $M(t,t',\kappa)$ respectively.
Furthermore, we denote by $M(0)$ the integral operator with the kernel
$$
M(t,t',0) :=
 C_M(\nu)\, \big(|W(t)| W(t')\, (1+t)(1+t')\big)^{\frac 12}\,
\left(\frac{1+t}{1+t'}\right)^{\nu\, {\rm sign}\,  (t-t')}
$$
where
$$
C_M(\nu) := -\frac{\pi}{2\sin(\nu\pi)\Gamma(1-\nu)\Gamma(1+\nu)}\, .
$$
Lemma \ref{HS-conv} in the Appendix says that $M(\kappa)$ converges in
the Hilbert-Schmidt norm to the operator $M(0)$ as $\kappa\to 0$, provided $W$
decays fast enough at infinity. This allows us to prove

\begin{Theorem} \label{firstThm}
Assume that $W$ satisfies \ref{firstassumption} and that
$\int_0^{\infty}\, (1+t)^{3-d}\, |W(t)|\, dt<\infty$, where $1\leq d< 2$.
Then the following statements hold true.
\begin{itemize}
\item[(a)]
If
$$
\int_0^{\infty}\, W(t)\, (1+t)^{d-1}\, dt < 0\, ,
$$
then the operator $B_0+\lambda\, W$ has at least one
negative eigenvalue for all $\lambda
>0$. For $\lambda$ small enough this eigenvalue, denoted by $E(\lambda)$,
is unique and satisfies
\begin{equation} \label{asymptotics-W}
\left(E(\lambda)\right)^{\frac{2-d}{2}}
 =  C(\nu)\left(\lambda\int_0^{\infty}\, W(t)\,
  (1+t)^{d-1}\, dt  +\mathcal{O}(\lambda^2) \right)\, ,
\end{equation}
where
$$
C(\nu) = \frac{\pi\, 2^{2\nu-1}}{(\Gamma(1-\nu))^2\,
  \sin(\nu\pi)} \, .
$$

\item[(b)]
If
$$
\int_0^{\infty}\, W(t)\, (1+t)^{d-1}\, dt > 0\, ,
$$
then the operator $B_0+\lambda\, W$ has no
negative eigenvalues for $\lambda$ positive and small enough.
\end{itemize}

\end{Theorem}

\begin{proof}
Part $(a)$.

\noindent The operator $B_0+\lambda\, W$ has eigenvalue $E=-\kappa^2$ if and
only if the operator
$$
\lambda K(\kappa) = \lambda M(\kappa)+\lambda L(\kappa)
$$
has an eigenvalue $-1$ for certain $\kappa(\lambda)$. On the other hand,
Lemma \ref{auxiliary} and (\ref{numbers}) imply that 
$$
 N_-\left(B_0+\lambda\,
  \frac{g_0}{g^+_0}\, V\right)\, \leq\, N_-(A_{\lambda,0})\, .
$$
The uniqueness of $E$, and so of $\kappa(\lambda)$, for $\lambda$ small enough
thus follows from (\ref{Bargman}) by taking $V= \frac{g^+_0}{g_0}\, W$.
Next we note that by Lemma \ref{HS-conv} 
for $\lambda$ small we have  $\lambda\, \|M(\kappa)\| < 1$ and
$$
\left(I+\lambda K(\kappa)\right)^{-1} = \left[I+\lambda(I+\lambda
  M(\kappa))^{-1}\, L(\kappa)\right]^{-1}\, (I+\lambda\, M(\kappa))^{-1}\, .
$$
So $\lambda K(\kappa)$ has an eigenvalue $-1$ if and only if
$\lambda(I+\lambda
  M(\kappa))^{-1}\, L(\kappa)$ has an eigenvalue $-1$. Since $\lambda(I+\lambda
  M(\kappa))^{-1}\, L(\kappa)$ is of rank one we get the equation for
  $\kappa(\lambda)$ in the form
\begin{equation} \label{kappa}
\tr\left(\lambda(I+\lambda
  M(\kappa(\lambda))^{-1}\, L(\kappa(\lambda)) \right) = - 1\, .
\end{equation}
Using the decomposition
$$
(I+\lambda M(\kappa))^{-1}= I -\lambda M(0)-\lambda (M(\kappa)-M(0)) +
\lambda^2 M^2(\kappa)(I+\lambda M(\kappa))^{-1}
$$
we obtain
\begin{eqnarray*}
&& \tr\left(\lambda(I+\lambda
  M(\kappa))^{-1}\, L(\kappa) \right)   \\
&& =
 \lambda\, C(\nu)\kappa^{-2\nu}\, \left(|W(t)|^{1/2}\, (1+t)^{-\nu+\frac
    12}\, , (I+\lambda M(\kappa))^{-1}W(t)^{1/2}\, (1+t)^{-\nu+\frac 12}
\right) \\
&& = C(\nu)\, \kappa^{-2\nu} \left(\lambda\, \int_0^{\infty}\, W(t)\,
  (1+t)^{d-1}\, dt +\mathcal{O}(\lambda^2)\right) \, .
\end{eqnarray*}
It thus follows from (\ref{kappa}) that
\begin{equation} \label{E}
E^{\nu}(\lambda) = -\kappa^{2\nu}(\lambda) =
C(\nu)\left(\lambda\int_0^{\infty}\, W(t)\,
  (1+t)^{d-1}\, dt  +\mathcal{O}(\lambda^2) \right)\, .
\end{equation}
To finish the proof of the part $(a)$ of the Theorem we mimic the argument
used in \cite{Si} and notice that if $(\varphi,(B_0+\lambda W)\, \varphi) <
0$, then $(\varphi, W\varphi) <0$, since $B_0$ is non-negative, and therefore
$(\varphi,(B_0+\tilde{\lambda} W)\, \varphi) < 0$ if $\lambda <
\tilde{\lambda}$. So if $B_0+\lambda W$ has a negative eigenvalue for
$\lambda$ small enough, then, by the variational principle,
it has at least one negative eigenvalue for all
$\lambda$ positive.

\noindent
Part $(b)$. From the proof of part $(a)$ it can be easily seen that if
$$
\int_0^{\infty}\, W(t)\, (1+t)^{d-1}\, dt > 0\, ,
$$
then $\tr\left(\lambda(I+\lambda
  M(\kappa))^{-1}\, L(\kappa) \right)$ is positive for $\lambda$ small and
therefore $K(\kappa)$ cannot have an eigenvalue $-1$.
\end{proof}

\begin{Remark}
Note that if
$$
W_0:= \int_{\R_+^2}\, W(t)\, W(t')(1+t)^{1-\nu}(1+t')^{1-\nu}
\left(\frac{1+t}{1+t'}\right)^{\nu\, {\rm sign}\,  (t-t')}\, dt\,
dt' <0\, ,
$$
then the operator $B_0+\lambda W$ has a negative eigenvalue for $\lambda$
small, positive or negative, also in the critical case when
$$
\int_0^{\infty}\, W(t)\, (1+t)^{d-1}\, dt = 0\, .
$$
Moreover, it follows from the proof of Theorem \ref{firstThm} that this
eigenvalue then satisfies
\begin{equation}
E^{\nu}(\lambda) =
C(\nu)\left(-\lambda^2\, C_M(\nu)\, W_0  +o\, (\lambda^2) \right)\, , \quad
\lambda\to 0\, .
\end{equation}

\end{Remark}

As an immediate consequence of Theorem \ref{firstThm} and inequalities
(\ref{eigenvalues}) we get

\begin{Theorem} \label{main}
Let $V$ satisfy assumption \ref{firstassumption} and let
$\int_0^{\infty}\, (1+t)^{3-d}\, |V(t)|\, dt<\infty$, where $1\leq d< 2$.
Then the following statements hold true.
\begin{itemize}
\item[(a)]
If
$$
\int_0^{\infty}\, V(t)\, g_0 (t)\, dt = \int_{\Gamma} V(|x|) \, dx < 0\, ,
$$
then the operator $A_{\lambda}$ has at least one
negative eigenvalue $E_{1,0}(\lambda)$ for all $\lambda
>0$. For $\lambda$ small enough this eigenvalue is unique and satisfies
\begin{equation} \label{inequality-V}
C_1\left|\lambda\int_{\Gamma}\, V(|x|)\,
\, dx \right|^{\frac{2}{2-d}} \,  \leq \, |E_{1,0}(\lambda)|\, 
\leq \, C_2 \left|\lambda\int_{\Gamma}\, V(|x|)\,
 dx \right|^{\frac{2}{2-d}}
\end{equation}
for suitable positive constants $C_1$ and $C_2$.

\item[(b)]
If
$$
\int_0^{\infty}\, V(t)\, g_0(t)\, dt = \int_{\Gamma} V(|x|) \, dx > 0\, ,
$$
then the discrete spectrum of $A_{\lambda}$ is empty
for $\lambda$ positive and small enough.
\end{itemize}
\end{Theorem}

\begin{proof}
Part (a). From (\ref{eigenvalues}) we get
$$
E^-_{1,0}(\lambda)   \leq E_{1,0}(\lambda) \leq
E^+_{1,0}(\lambda) \, .
$$
Moreover, by Lemma \ref{auxiliary} $E^{\pm}_{1,0}(\lambda)$ are the lowest
eigenvalues of operators $B_{\lambda,0}^{\pm}$. The existence and uniqueness
of $E_{1,0}$ thus follows from part (a) of Theorem \ref{firstThm} applied with
$W(t)= V_0^+(t)$ and $W(t)= V_0^-(t)$ respectively. At the same time,  
equation (\ref{asymptotics-W}) implies (\ref{inequality-V}).

Similarly, part (b) of the statement follows immediately from Lemma
\ref{auxiliary} and part (b) of Theorem \ref{firstThm} applied with  $W(t)=
V_0^-(t)$.
\end{proof}

\begin{Remark}
We note that the strong coupling behavior of $A_{\lambda}$ is, on the
contrary to (\ref{inequality-V}), typically one-dimensional, 
i.e. determined by the local dimension of $\Gamma$. 
Namely, if $V$ is continuous and compactly supported,
then the standard Dirichlet-Neumann bracketing technique shows
that the Weyl asymptotic formula 
$$
\lim_{\lambda\to\infty}\, \lambda^{-\gamma-\frac 12}\, \sum_j\, |E_j|^{\gamma}
=  L_{\gamma,1}^{cl}  
\, \int_{\Gamma}\, |V|^{\gamma+\frac 12}\, dx, \quad \gamma \geq 0
$$
holds true, where $E_j$ are the negative eigenvalues of $A_{\lambda}$ and
$L_{\gamma,1}^{cl}= \frac{\Gamma(\gamma+1)}{2\sqrt{\pi}\,
  \Gamma(\gamma+3/2)}$. 
\end{Remark}

\subsection{The case $d=2$}
For $d=2$ one can mimic the above procedure replacing the Hankel functions
$H_{\nu}^{(1,2)}$ by $H_0^{(1,2)}$. The latter have a logarithmic
singularity at zero and therefore  it turns out that
the lowest eigenvalue of $A_{\lambda}$ then converges to
zero exponentially fast. Indeed, here instead of
(\ref{asymptotics-W}) one obtains
$$
E(\lambda) \sim -e^{-\lambda^{-1}}\, ,
$$
as for the two-dimensional Schr\"odinger operator, see \cite{Si}. Since
the analysis of this case is completely analogous to the previous one,
we skip it.

\subsection{The case $d>2$}
\label{virtual}

Now it remains to show that for $d>2$ and $\lambda$ small enough 
the discrete spectrum of $A_{\lambda}$, which in this case 
coincides with that of $A_{\lambda,0}$, remains empty. Since the discrete
spectrum of $A_{\lambda,0}$ might only consist of negative eigenvalues, see
(\ref{essential-k}), it suffices to prove  

\begin{Proposition} \label{virtual-level}
If $d>2$ and $V\in L^{\infty}(\R_+)\cap L^{d/2}(\R_+,g_0)$,
then there exists $\lambda_0>0$ such that
the operator $A_{\lambda,0}$ is non-negative for all $\lambda\in
[0,\lambda_0]$.
\end{Proposition}

\begin{proof}

Consider a function $f\in D(Q_0)$. Since $f\in H^1(\R_+)$, which is
continuously embedded in $L^{\infty}(\R_+)$, it follows that $f\to 0$
at infinity and we can write
$$
f(t) = -\int_t^{\infty}\, f'(s)\, ds\, .
$$
In view of (\ref{gkestim}) we have $g_0^{-1}\in L^1(\R_+)$. 
Using Cauchy-Schwarz inequality we thus find out that for any $q\geq q_0$, 
where
$
\frac{1}{q_0}+\frac 1d=\frac 12\, ,
$
the following estimate holds true 
\begin{eqnarray} \label{sobolev}
& & \left(\int_0^{\infty} |f(t)|^q\, g_0(t)\, dt\right)^{\frac 1q}  \leq 
\left(\int_0^{\infty}\left( \int_t^{\infty}\, 
|f'(s)| ds\right)^q\, g_0(t)\, dt\right)^{\frac 1q} \nonumber \\
& & \quad \leq 
\left(\int_0^{\infty}\left( \int_t^{\infty}\, 
|f'(s)|^2 g_0(s)\, ds\right)^{\frac q2}\, \left(\int_t^{\infty}\,
\frac{ds}{g_0(s)}\right)^{\frac q2}   
\, g_0(t)\, dt\right)^{\frac 1q} \nonumber \\
&& \quad \leq  C(q)\,  \left(\int_0^{\infty} |f'(s)|^2\, g_0(s)\,
  ds\right)^{\frac 12}\, ,
\end{eqnarray}
with a constant $C(q)$ independent of $f$. 
The H\"older inequality and
(\ref{sobolev}) then give
\begin{eqnarray*}
\int_0^{\infty}\, |V|\, |f|^2 g_0(t)\, dt  & \leq &
\left(\int_0^{\infty} |V|^{d/2}\, g_0(t)\, dt\right)^{\frac 2d}
\left(\int_0^{\infty} |f|^{q_0}\, g_0(t)\, dt\right)^{\frac{2}{q_0}} \nonumber
\\
& \leq  & C^2(q_0)\, \int_0^{\infty} |f'|^2\, g_0(t)\, dt\,
\left(\int_0^{\infty} |V|^{d/2}\, g_0(t)\, dt\right)^{\frac 2d}\nonumber \, ,
\end{eqnarray*}
which implies
$$
Q_0[f] \geq \int_0^{\infty} |f'|^2\, g_0(t)\, dt\left [1- \lambda\, C^2(q_0)
\left(\int_0^{\infty} |V|^{d/2}\, g_0(t)\, dt\right)^{\frac 2d}\right ]\, .
$$
To finish the proof it suffices to take $\lambda$ small enough so that
$Q_0[f]\geq 0$.
\end{proof}

\section*{Appendix}

\begin{Lemma} \label{HS-conv}
Let $W$ be bounded and assume that $\int_0^{\infty}\, (1+t)^{1+2\nu}\,
|W(t)|\, dt<\infty$. Then $M(\kappa)$ converges in the
Hilbert-Schmidt norm to the operator $M(0)$ as $\kappa\to 0$.
\end{Lemma}

\begin{proof}
We first notice that $M(0)$ is Hilbert-Schmidt, since
$$
\int_0^{\infty}\int_0^{\infty}\, |M(t,t',0)|^2\, dt\, dt' < \infty
$$
by assumption. We will also need the asymptotic behavior of the Bessel
functions with purely imaginary argument near zero:
\begin{equation} \label{J-asymp}
J_{\nu}(i\kappa(1+t)) = e^{i\pi\nu /2}\, I_{\nu}(\kappa(1+t)) \sim
e^{i\pi\nu /2}\, \frac{\kappa^{\nu}(1+t)^{\nu}}
{2^{\nu}\,\Gamma(\nu+1)}\, ,\quad
\kappa(1+t)\to 0\, ,
\end{equation}
see \cite{AS,W}. From the definition of Hankel's functions we thus get
$$
\beta(\kappa)= \frac{J_{1-\nu}(i\kappa)-e^{i(1-\nu)\pi}\,
  J_{\nu-1}(i\kappa)}{e^{i(\nu-1)\pi}\, J_{\nu-1}(i\kappa)-J_{1-\nu}(i\kappa)}
\, \to -e^{-2i\nu\pi}\, ,\quad \kappa\to 0\, .
$$
This together with the asymptotics (\ref{J-asymp}) implies
\begin{equation} \label{K-asymp}
\lim_{\kappa\to 0}\, M(t,t',\kappa)= M(t,t',0)\,
.
\end{equation}
Now using the asymptotic behavior of Hankel's functions at infinity,
\cite{AS}, we find out that
$$
G(t,t',\kappa) \sim \left((1+t)(1+t')\right)^{1/2}\,
\frac{e^{-\kappa(t+t')}-\beta(\kappa)\, e^{-\kappa|t-t'|}}{\kappa\,
    (1+t)(1+t')^{1/2}}\, , \quad \kappa^2(1+t)(1+t')\to \infty\, .
$$
Since $|\beta(\kappa)|$ is bounded,
we obtain the following estimates.

\noindent For $\kappa^2(1+t)(1+t')\geq 1$:
$$
|K(t,t',\kappa)|\, , |L(t,t',\kappa)| \leq C\, \left|W(t')\, W(t)
  (1+t)(1+t')\right|^{1/2}\, .
$$
For $\kappa^2(1+t)(1+t')< 1$:
$$
|M(t,t',\kappa)| \leq C'\, |W(t')\, W(t) \, |
\left [1+((1+t)(1+t'))^{\nu+\frac 12}\right]\, ,
$$
where  we have used (\ref{K-asymp}). Note that the
constants $C$ and $C'$ may be chosen independent of $\kappa$,
which enables us to employ
the Lebesgue dominated convergence theorem to conclude that
$$
\lim_{\kappa\to 0}\, \int_{\R^2_+}\, |M(t,t',\kappa)-
M(t,t',0)|^2\, dt\, dt' = 0\, .
$$
\end{proof}

\section*{Acknowledgement}
%
The work has been partially supported by
the Czech Academy of Sciences and by DAAD within 
the project D-CZ~5/05-06.

%
%
\providecommand{\bysame}{\leavevmode\hbox to3em{\hrulefill}\thinspace}
\providecommand{\MR}{\relax\ifhmode\unskip\space\fi MR }
\providecommand{\MRhref}[2]{%
  \href{http://www.ams.org/mathscinet-getitem?mr=#1}{#2}
}
\providecommand{\href}[2]{#2}
\end{document}